\documentstyle[floats,twocolumn,aps,prl,epsf,rotate] {revtex}

\begin{document}

\draft
\tighten

\title{Poisoning of Hydrogen Dissociation at Pd (100) by Adsorbed Sulfur
Studied by {\em ab initio} Quantum Dynamics and  {\em ab initio}
Molecular Dynamics}

\author{Axel Gross\cite{elm}, Ching-Ming Wei\cite{Wei_add}, 
and Matthias Scheffler}

\address{Fritz-Haber-Institut der Max-Planck-Gesellschaft, Faradayweg 4-6, 
D-14195 Berlin-Dahlem, Germany
}

\maketitle

\begin{abstract}

We report calculations of the dissociative adsorption of H$_2$ at 
Pd (100) covered with 1/4 monolayer of sulfur using quantum dynamics as well as
molecular dynamics and taking all six degrees of freedom of the two H
atoms fully into account. The {\em ab initio} potential-energy surface (PES) 
is found to be very strongly corrugated. In particular we discuss the
influence of tunneling, zero-point vibrations,
localization of the nuclei's wave function when narrow
valleys of the PES are passed,
steering of the approaching H$_2$
molecules towards low energy barrier configurations, and the time
scales of the center of mass motion and the other degrees of freedom.
Several ``established'' concepts, which were derived from
low-dimensional dynamical studies, are shown to be not valid.

\end{abstract}


The presence of an adsorbate on a surface can profoundly change the 
surface reactivity. An understanding of the underlying mechanisms
is of relevance for, e.g., corrosion, lubrication, and catalysis. 
Typically the breaking of
molecular bonds is the rate-limiting step in a 
surface chemical reaction, and the model system 
is the dissociation and adsorption  of hydrogen.
Whereas on clean Pd (100) the dissociation of H$_2$ happens very
efficiently, it is well known that a small amount of adsorbed sulfur 
reduces (poisons) the surface reactivity significantly \cite{Ren89,Bur90}.
Density functional theory (DFT) calculations have shown
that hydrogen dissociation on sulfur-covered Pd(100) is still exothermic,
however, the dissociation is hindered by the formation of energy barriers
in the entrance channel of the potential-energy surface (PES) \cite{Wil96PRL}.

We have recently extended this study and determined the 
six-dimensional PES of the system H$_2$/S(2$\times$2)/Pd(100) 
in great detail \cite{Wei97} using DFT together 
with the generalized  gradient approximation (GGA) \cite{Per92}.
On an analytical representation of this {\it ab initio} PES we have 
now performed six-dimensional quantum and classical dynamics
calculations in which all hydrogen degrees of freedom are treated dynamically. 
The methods are  described in detail in Ref.~\cite{Gro98PRB}.
The study is based essentially on the following approximations: ($i$) 
Born-Oppenheimer approximation,
($ii$) the substrate is kept rigid
($iii$) the exchange-correlation functional is treated in the GGA, 
($iv$) the DFT-GGA calculations are performed in the supercell approach,
($v$) the {\it ab initio} PES is fitted to an analytical form,

Items $i)$ and $ii)$ mean that energy dissipation is not treated 
{\em explicitly}. This is appropriate because, while energy dissipation 
plays a crucial role by taking away the {\em adsorption} energy, for the 
{\em dissociation} of H$_2$ energy dissipation into substrate or electronic
degrees of freedom is not significant. The reason lies in the
involved time scales and the mass differences between 
hydrogen and S and Pd substrate atoms.

The GGA treatment of the exchange-correlation functional [item $iii)$]
is the best treatment known to date. It leaves an uncertainty to
total-energy differences typically below 0.1 eV (see, e.g., Ref.~\cite{Ham94}).
The error due to approximation $iv)$ was checked to be smaller than 0.1 eV
\cite{Wil96PRL,Wei97}. 
The fit of the PES [item $v)$] by an analytic function
has been performed in such a way that the relative error between the fit
and the original data, $|\Delta E|/|E|$, is smaller than 5 \%.
For further details of the DFT calculations and the analytical 
representation of the PES see Ref.~\cite{Wei97}. We note
that the PES of this system is very corrugated, which represents
a significant challenge for a proper
treatment of the quantum dynamics.
Even earlier empirical studies had not considered such corrugations
(energy barriers between 0.09 eV for the optimum pathway and 2.5 eV 
for an approach over the S adatom \cite{Wei97}). Nevertheless, we 
ensured that the error due to the finite number of channels considered in 
the quantum dynamical calculations is below 2 \% of the sticking
probability.

The corrugation of the PES has several unexpected consequences.
For example,  it leads to  zero-point vibrations (ZPV) in the quantum 
dynamics which reduce the quantum 
sticking probabilities compared to the classical results. Since, however, the
ZPV energies in the single frustrated modes are rather
small, the quantum particles do not realize  completely the ZPV
in these slow modes. Furthermore we find that
steering of molecules to low-barrier configurations is effective
at unusually high kinetic energies.
And in spite of the fact that the PES exhibits a early minimum barrier
for dissociation, initial vibrational excitation of impinging
molecules enhances the dissociation probability. These results
confirm that results derived from statical analyses of the PES,
as e.g. the barrier distribution, are not sufficient to obtain
reaction probabilities, but a proper {\em high-dimensional}
{\em  dynamical}
treatment of the dissociation process is essential.

\begin{figure}[tb]
\unitlength1cm
\begin{center}
   \begin{picture}(10,7.0)
\centerline{   {\epsfxsize=8.cm  
          \epsffile{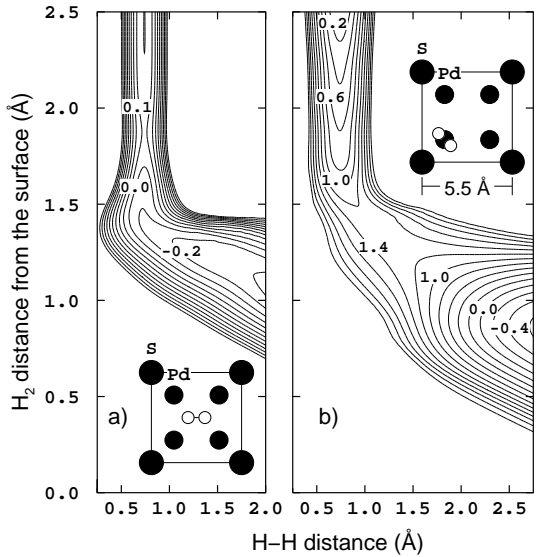}} }
   \end{picture}
\end{center}

\vspace{-1.cm}

   \caption{Lines of constant energy for two cuts 
            through the six-dimensional PES of 
            H$_2$ at S(2$\times$2)/Pd\,(100). The insets show the
            orientation of the molecular axis and the lateral
            H$_2$ center-of-mass coordinates, i.e. the coordinates
            $X$, $Y$, $\theta$, and $\phi$. The coordinates 
            in the figure are the H$_2$ center-of-mass distance 
            from the surface $Z$ and the H-H interatomic distance $r$. 
            Energies are in eV per H$_2$ molecule.
            The contour spacing in Fig.~\protect\ref{elbow}a  is 0.1~eV,
            while in Fig.~\protect\ref{elbow}b it is 0.2~eV.
            Fig.~\protect\ref{elbow}a corresponds to the minimum energy
            pathway.}

\label{elbow}
\end{figure}

Figure~\ref{elbow} shows two 
cuts through the six-dimensional PES with the optimum pathway
displayed in Fig.~\ref{elbow}a. The 
minimum energy barrier to dissociation ($E_{\rm b} =$~0.09~eV)
lies at a configuration where the bond-length of the molecule is
is nearly unchanged compared to the gas phase 
value \cite{Wil96PRL,Wei97}. In such a situation it is usually anticipated
that the vibrational and translational degrees of freedom are almost
uncoupled so that vibrational energy of the impinging molecules cannot
be used to overcome the barrier (see, e.g., Ref.~\cite{Zan88,Dar95}).
Consequently, it has been predicted \cite{Wil96PRL} that the sticking 
probability of H$_2$ at S(2$\times$2)/Pd(100) should show no strong 
dependence on the initial vibrational state of the molecule. 
This prediction corresponds to the so-called Polanyi rules which have 
been formulated for gas-phase dynamics thirty years ago \cite{Pol69}.
We will show that these rules have to be modified for strongly
corrugated PESs.

The large repulsion for hydrogen dissociation closer to the
sulfur atoms is illustrated in Fig.~\ref{elbow}b. While the dissociation 
path over the on-top position on the clean surface is hindered by a 
barrier of height 0.15~eV \cite{Wil96PRB}, the adsorbed sulfur leads 
to an increase in this barrier height to 1.3~eV.

\begin{figure}[tb]
\unitlength1cm
\begin{center}
   \begin{picture}(10,4.5)
\centerline{   \rotate[r]{\epsfysize=8.cm  
          \epsffile{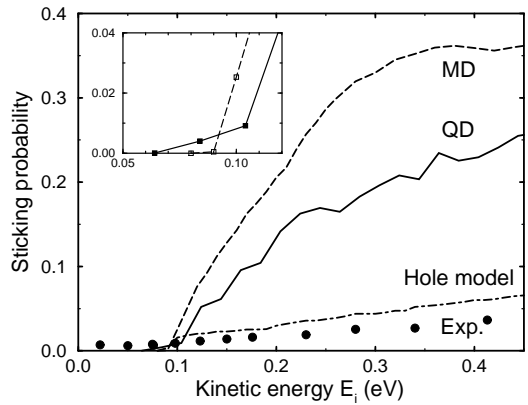}} }
   \end{picture}

\end{center}
   \caption{Sticking probability versus kinetic energy for
a H$_2$ beam under normal incidence on a S(2$\times$2)/Pd(100) surface.
Full dots: experiment (from ref.~\protect\cite{Ren89});
dashed-dotted line: integrated barrier distribution,
which corresponds to the sticking probability in the hole 
model \protect\cite{Kar87};
solid line: quantum mechanical results for molecules initially in the
rotational and vibrational ground-state (QD);
dashed line: classical molecular dynamics results for initially 
non-rotating and non-vibrating molecules (MD). 
The inset shows the quantum and classical results at low 
energies.}

\label{stick}
\end{figure}

Figure \ref{stick} compares our results for the sticking 
probability as a function of the kinetic energy of the 
incident H$_2$ beam with the experiment \cite{Ren89}.
In addition, also the integrated barrier distribution
$P_b(E)$ is plotted, which is the fraction
of the configuration space for which the barrier towards dissociation
is less than $E$ and which is also called the ``hole model'' \cite{Kar87}.

The calculated sticking probabilities are significantly larger than the
experimental results.
The agreement between the adsorption experiment and our calculations
is limited to the single aspect that adsorbed sulfur is a strong poison
for the H$_2$ dissociation, and the onset energy for dissociative adsorption, calculated as $E_{\rm i} \approx 0.12$~eV, is in fact in excellent
agreement with the experimentally measured mean kinetic energy of
hydrogen molecules desorbing from sulfur covered Pd(100) \cite{Com80}. 
Otherwise, in the energy range $E_{\rm i} > 0.12 $ eV,
the calculated sticking probabilities are significantly larger
than the experimental results.
We consider this a most important result, because progress in science 
is typically achieved by disagreement between theoretical and experimental
results rather than by agreement. What are the likely reasons for the 
accounted disagreement?

$i)$ From the approximations assumed in the calculations (see above)
it is mainly the GGA which bears the uncertainty to
be good enough for a reliable description of the {\em topology} of
the PES. It should be noted that $S(E_{\rm i})$ probes the details of
the energy corrugation of the PES and not just the lowest energy barrier.
We note, however,  that our results are not very sensitive to
an uncertainty of the PES of $< 0.2$ eV, which we tested by adding
some artificial functions to our {\em ab initio} PES,
and that there is no indication in the literature (so far) that the
GGA would not be able to deal with such a required, modest accuracy.

$ii)$ The other approximations have been carefully
tested and we do not believe that they are relevant.

$iii)$ Concerning the experimental results we note that the
determination of the coverage is not without problems.  In fact,
it is well possible that the adsorbate coverage in the measurements 
was higher than reported.  The sulfur adlayer was obtained by
simply heating the sample which leads to the segregation of 
bulk sulfur at the surface. The coverage of $\Theta_{\rm S} = 0.25$
was then identified by monitoring the S$_{132}$ and Pd$_{330}$ Auger 
signals with respect to those at saturation coverage which is believed 
to be $\Theta_{\rm S} = 0.5$, and by the presence of a $(2 \times 2)$
LEED pattern. 
We speculate that the actual coverage at which the data of Fig.
\ref{stick} were taken could well have been higher, e.g. due to some
significant amount of subsurface sulfur. Subsurface adsorbates
can affect the surface reactivity significantly.
We hope that more experiments of the type
done by Rendulic {\em et al.} \cite{Ren89} will be performed and that
also other techniques for producing the S adlayer will be considered,
in order to determine in what form S is actually present at the surface
under different experimental conditions.

We now turn back to the theoretical results. The classical treatment
of hydrogen  dynamics over-estimate the sticking probability 
of H$_2$ at S(2$\times$2)/Pd(100) compared to the quantum results. 
For energies smaller than the minimum barrier height $E_{\rm b}$ the classical
sticking probability is of course zero, whereas the quantum results
still show some dissociation due to tunneling, as the inset of 
Fig.~\ref{stick} reveals. But for energies larger than
$E_{\rm b}$ the classical sticking probability rises to 
values which are almost 50\% larger than the quantum sticking
probabilities.
We have recently shown that ZPV can lead to
a suppression of the quantum sticking probability compared to classical
results \cite{Gro98PRB,Gro97Vac}. 
In Tab.~\ref{zeros} we have therefore
collected the ZPV energies of the hydrogen molecule at the
minimum barrier position.
Since here the hydrogen molecular bond is still almost intact,
the zero-point energy in the H-H vibration has only decreased 
by 5~meV from the gas-phase ZPV energy of H$_2$. 
However, along the minimum energy path the molecule has to pass through a 
narrow valley of the PES in the lateral coordinates parallel to the surface
due to the strong corrugation of the PES. This causes substantial
ZPV energies in the frustrated lateral modes of the molecular 
center-of-mass of motion already relatively far away from the surface. 
Note that the ZPV energies in the frustrated rotational modes 
are fairly small in comparison. This is again due to the
fact that the molecular bond is essentially not elongated at the minimum
barrier position so that the molecular interaction with the surface
is still rather isotropic.

\begin{table}
\begin{center}
\begin{tabular}{|c|l|}
mode & ZPV (eV)\\
\hline
H-H vibration & 0.253 \\
polar rotation & 0.016 \\
azimuthal rotation & 0.013 \\
translation perpendicular to molecular axis & 0.027 \\
translation parallel to molecular axis & 0.027 \\
\hline
sum & 0.336 \\
\end{tabular}
\caption{Zero-point vibrations (ZPV) energies of the H$_2$ molecule at the
minimum barrier position. The H$_2$ configuration
corresponds to the situation of Fig.~\protect\ref{elbow}a. 
The gas-phase ZPV energy of H$_2$ is
$\frac{1}{2} \hbar \omega_{\rm gas} =$~0.258~eV.
}
\label{zeros}
\end{center}

\vspace{-1.2cm}

\end{table}

Because the sum of all ZPV energies at the minimum barrier position
is  0.08~eV larger than the gas-phase ZPV energy of
hydrogen, the effective minimum barrier height for the quantum particles 
should be increased by 0.08~eV to $E_{\rm b}^{\rm eff} = $~0.17~eV. 
And indeed the quantum
sticking probabilities are smaller than the classical sticking probabilities 
for energies larger than the classical minimum barrier height. 
The absence of ZPV in the classical dynamics leads to the 
enhanced sticking probability in the classical calculations \cite{Gro97Vac}.
However, the quantum sticking probability rises significantly already
for energies well below the {\em effective} barrier height of 0.17 eV. 
This can be understood by considering the
involved time-scales. A vibration with $\hbar \omega = $~50~meV, e.g., 
corresponds to a vibrational period of 80 fs. 
The time it takes for an H$_2$ molecule with $E_{\rm i} = 0.12$ eV 
to pass the barrier (width 0.5 \AA) is only 25 fs.
This means that during the crossing of the minimum barrier
region the hydrogen molecule would hardly perform one complete vibration
in any of the four frustrated modes given in Tab.~\ref{zeros}. 
The concept that in the absence of tunneling a quantum particle needs
at least the barrier energy plus the ZPV energies in order to
cross a barrier is evidently not applicable for such slow modes.
The quantum particles do not completely follow the zero-point motion 
perpendicular to the reaction path and thus do not additionally need 
the total ZPV energies in order to cross the barrier.

The question arises why the difference between classical and quantum 
sticking probabilities even increases with increasing energy
although ZPVs should become less important at higher kinetic energies. 
An analysis of the ZPV shows that the sum of the four ZPV energies
of the frustrated lateral and rotational modes actually rises to values 
of 0.2~eV along the reaction pathways. Although the H-H vibrational energy
further decreases, the sum of {\em all} five ZPV energies becomes even larger 
than the value of 0.336~eV at the minimum barrier position.
This sum plays a crucial role in the quantum 
dynamics not only at the minimum barrier position, but also later along 
the dissociation pathways. Since at different kinetic energies different
parts of the barrier region with varying ZPV energies are
sampled, classical and quantum results are not just shifted by a fixed
amount, and the ZPV need to be included in the full
dynamical treatmemt.

A comparison of the dynamical results with the integrated barrier 
distribution in Fig.~\ref{stick} shows that the sticking probabilities are
much larger than what one would expect from the hole model \cite{Kar87}.
Our analysis of swarms of trajectories with different initial conditions
shows that molecules directed towards high barriers are very efficiently 
steered to configurations and sites with lower barriers \cite{Gro95PRL,Kay95}. 
Indeed the strongly repulsive potential above the sulfur atoms 
extends rather far into the gas-phase \cite{Wei97}. Thus the forces
re-orienting and re-directing the molecules act on the molecules for
a long time; therefore they are so effective up to kinetic energies
of more than 0.5~eV. At the clean Pd surface,
on the contrary, at $E_{\rm i} \approx 0.2$~eV
steering is no longer operative in the dissociation process \cite{Gro95PRL}. 
This causes the surprising result that at $E_{\rm i} = 0.2$~eV the
sticking probability at the  (2$\times$2) sulfur-covered Pd(100) 
surface is almost as large as at the clean Pd(100) surface in spite 
of the fact that at the clean surface the fraction of open
dissociation pathways is more than five times larger at this energy
\cite{GroCPL96a}.

\begin{figure}[tb]
\unitlength1cm
\begin{center}
   \begin{picture}(10,4.5)
\centerline{   \rotate[r]{\epsfysize=8.cm  
          \epsffile{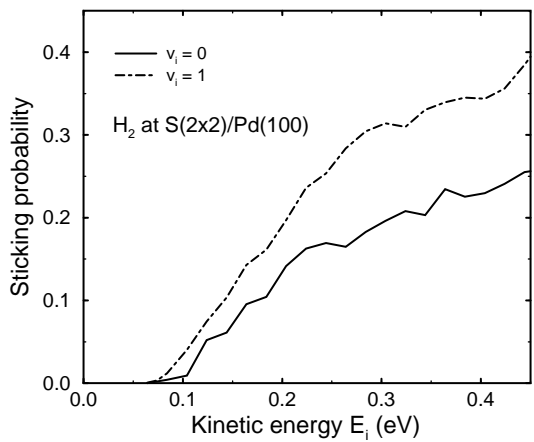}} }
   \end{picture}

\end{center}
   \caption{Dependence of the quantum sticking probability versus 
kinetic energy for a hydrogen beam under normal incidence at a 
S(2$\times$2)/Pd(100) surface on the initial vibrational state $v$.
Solid line: $v_i = 0$, dash-dotted line: $v_i = 1$. The molecules
are initially in the rotational ground state.}
\label{state_stick}
\end{figure}

The effect of initial vibrational motion on the dissociation
process is illustrated in Fig.~\ref{state_stick} where state-specific
sticking probabilities as a function of the incident kinetic energy are 
plotted. As noted above, since the PES has an early minimum barrier for
dissociation, no coupling of the initial vibration to the dissociation
is expected. However, as Fig.~\ref{state_stick} demonstrates, initial
vibrational excitation leads to a significant increase in the sticking 
probability. This shows that the dissociation process cannot be understood
by simply analyzing the minimum energy path. As Fig.~\ref{elbow}b reveals,
there are molecular configurations for which the bond is significantly
extended at the barrier position. At higher kinetic energies
the adsorbing molecules also probe such pathways along which the
vibrational energy can be efficiently used to overcome
the dissociation barrier.
The unexpected result of the vibrationally enhanced dissociation in spite
of an early minimum barrier to dissociation is actually in agreement
with the experimentally observed vibrational over-population in 
thermal hydrogen desorption from sulfur covered Pd(100) \cite{SchDiss91}
invoking the principle of microscopic reversibility.
We have also studied the dependence of the sticking probability on
the initial rotational quantum number. The rotational effects are
similiar to the ones found at the clean surface and will be discussed in a
forthcoming publication.

In conclusion, we reported a six-dimensional dynamical
study of the dissociative adsorption of H$_2$ at S(2$\times$2)/Pd(100)
employing a PES obtained from detailed density 
functional theory calculations. The dynamical results reproduce the 
poisoning effect of sulfur adsorption for hydrogen dissociation on Pd(100),
but large quantitive differences to the experiment exist. 
The huge corrugation and anisotropy 
of the PES lead to ZPV in the 
quantum dynamics which reduce the quantum results compared to the classical
ones. However, at the minumum barrier the additional ZPV
correspond to very slow modes so that the quantum particles do not completely
follow these modes. In addition, the huge corrugation and anisotropy
cause large steering effects at unusual high kinetic energies. 
The multi-dimensionality of the PES furthermore leads to an enhancement
of the sticking probability for vibrating molecules although
the PES exhibits an early barrier for dissociative adsorption.

\end{document}